\documentclass[pre,aps,twocolumn]{revtex4}

\usepackage{graphicx}
\usepackage{amsmath}
\usepackage{amssymb}
\usepackage{ifthen}
\usepackage{booktabs}

\usepackage{graphicx}
\usepackage{dcolumn}
\usepackage{bm}
\usepackage{longtable}
\usepackage{color}

\newcommand{\bb}{\begin{equation}}
\newcommand{\ee}{\end{equation}}
\newcommand{\ba}{\begin{eqnarray*}}
\newcommand{\ea}{\end{eqnarray*}}

\begin{document}

\title{Crossover scaling of apparent first-order wetting in two dimensional systems with short-ranged forces}

\author{Andrew O. \surname{Parry}}
\affiliation{Department of Mathematics, Imperial College London, London SW7 2B7, UK}
\author{Alexandr \surname{Malijevsk\'y}}
\affiliation{
{Department of Physical Chemistry, University of Chemical Technology Prague, Praha 6, 166 28, Czech Republic;}\\
 {Institute of Chemical Process Fundamentals of the Czech Academy of Sciences, v. v. i., 165 02 Prague 6, Czech Republic}}

\begin{abstract}
Recent analyses of wetting in the semi-infinite two dimensional Ising model, extended to  include both a surface coupling enhancement and a surface
field, have shown that the wetting transition may be effectively first-order and that surprisingly the surface susceptibility develops a divergence
described by an anomalous exponent with value $\gamma_{11}^{\rm eff}=\frac{3}{2}$. We reproduce these results using an interfacial Hamiltonian model
making connection with previous studies of two dimensional wetting and show that they follow from the simple crossover scaling of the singular
contribution to the surface free-energy which describes the change from apparent first-order to continuous (critical) wetting due to interfacial
tunnelling. The crossover scaling functions are calculated explicitly within both the strong-fluctuation and intermediate-fluctuation regimes and
determine uniquely and more generally the value of $\gamma_{11}^{\rm eff}$ which is non-universal for the latter regime. The location and the
rounding of a line of pseudo pre-wetting transitions occurring above the wetting temperature and off bulk coexistence, together with the crossover
scaling of the parallel correlation length, is also discussed in detail.
\end{abstract}

\maketitle

\section{Introduction}
Abraham's exact solution of the semi-infinite planar Ising model showed a  wetting transition which was continuous and strictly second-order i.e. the
surface specific heat exponent takes the value $\alpha_s=0$ \cite{Abraham1980}. Subsequent studies based on interfacial Hamiltonian models, and also
random walk arguments gave strong support that this is the general result for 2D wetting in systems with short ranged forces and describes a
universality class, referred to as the strong-fluctuation (SFL) regime \cite{Fisher,Schick,Forgacs}. In particular renormalization group analyses of
interfacial models show that for systems with strictly short-ranged forces the flow is described by only two non-trivial fixed points describing a
bound phase (characterising the SFL regime) and an unbound phase respectively \cite{LipowskyFisher1,Huse}. While first-order wetting transitions are
possible in 2D they require the presence of sufficiently long-ranged intermolecular forces \cite{ZLK,Privman,LN}. However, very recently exact and
numerical studies of the wetting transition in the Ising model, but now including an additional short-ranged field representing the enhancement of
the surface coupling constant, have shown that the wetting transition is effectively first-order when the coupling constant is large
\cite{Abraham2016}.This enhancement of the surface coupling, which acts  in addition to a surface field, is similar to the well known mechanism which
drives wetting transitions first-order in mean-field treatments of Ising and lattice-gas models \cite{sullivan}. What is most surprising here is that
it was observed that on approaching the wetting temperature the surface susceptibility and specific heat appear to diverge and are characterised by
an anomalous exponent equal to $3/2$ before saturating to a very large finite value. In this paper we place these results within the more general
theory of 2D wetting based on interfacial Hamiltonians and show that they are consistent with a simple scaling theory for the crossover from apparent
first-order to critical wetting within both the SFL and intermediate fluctuation scaling regimes - these are the regimes in which the interface has
to tunnel through a potential barrier in order to unbind from the wall. We also discuss the location and rounding of a line of pseudo pre-wetting
transitions occurring above the wetting temperature which serves to emphasise the effective first-order nature of the wetting transition.

\section{Scaling and fluctuation regimes for 2D critical wetting}
{\bf Background:} The fluctuation theory of wetting transitions, particularly those occurring  in 2D systems, was successfully developed several
decades ago; see for example the excellent and comprehensive review articles \cite{Fisher,Schick,Forgacs}. Wetting transitions refer to the change
from partial wetting (finite contact angle) to complete wetting (zero contact angle) which occurs at a wetting temperature $T_w$. Viewed in the grand
canonical ensemble the wetting transition, occurring at a wall-gas interface say, is associated with the change from microscopic to macroscopic
adsorption of liquid as $T\to T_w^-$ at bulk coexistence. The transition is therefore equivalent to the unbinding of the liquid-gas interface, whose
thermal fluctuations are resisted by the surface tension $\sigma$. The transition may be first-order or continuous (often termed critical wetting) as
identified from the vanishing of the singular contribution to the wall-gas surface tension $\sigma_{\rm sing}\equiv\sigma(\cos\theta-1)\propto
-(T_w-T)^{2-\alpha_s}$. Thus in standard Ehrenfest classification the value $\alpha_s=1$ corresponds to first-order wetting and is usually associated
with the abrupt divergence of the equilibrium adsorption (proportional to the wetting film thickness $\langle\ell\rangle$) as $T\to T_w^-$. In 3D the
transition is also associated with a pre-wetting line of thin-thick transitions extending above $T_w$ and off coexistence which terminates at a
pre-wetting critical point. For critical wetting the exponent $\alpha_s<1$ and we need to introduce further critical exponents for the film
thickness, $\langle\ell\rangle\propto (T_w-T)^{-\beta_s}$, and parallel correlation length, $\xi_\parallel\propto (T_w-T)^{-\nu_\parallel}$, which
diverge continuously on approaching the transition. In the near vicinity of the transition, the free-energy shows scaling $\sigma_{\rm sing}=
t^{2-\alpha_s}W(h t^{-\Delta_s})$ where $t\propto (T_w-T)$ and $h$ (measuring the bulk ordering field or deviation from liquid-gas coexistence) are
the two relevant scaling fields for critical wetting. Here $W(x)$ is a scaling function, $\Delta_s$ is the surface gap exponent and we have
suppressed metric factors for the moment. As is well known the scaling of the free-energy is a powerful constraint on the critical singularities. For
example it follows that the exponents satisfy standard relations such as the Rushbrooke-like equality $2-\alpha_s=2\nu_\parallel-2\beta_s$. With the
additional assumption of hyperscaling, which in 2D implies $2-\alpha_s=\nu_\parallel$ the gap exponent follows as $\Delta_s=3\nu_\parallel/2$ leaving
just one exponent undetermined. Random walk arguments go further and for short-ranged forces determine uniquely the values of the critical
singularities at critical wetting in terms of the interfacial wandering exponent for a free interface \cite{Fisher}. For pure systems with thermal
disorder this determines, $\alpha_s=0$, $\beta_s=1$ and $\nu_\parallel=2$ (and hence $\Delta_s=3$) in keeping  with Abraham's exact Ising model
results. More recently, studies of  fluid adsorption in other geometries, in particular wedge filling, have revealed a number of unexpected geometry
invariant properties of wetting \cite{wood} whose microscopic origins have been illuminated by very powerful field theoretic formulations of phase
separation \cite{delfino}. Finally we note that scaling theories pertinent to first-order wetting transitions have also been developed and been used
in particular to analyse the critical singularities associated with the line tension \cite{IR,I}. We shall return to this later.

These remarks are completely supported by analyses of wetting based on interfacial  Hamiltonians which have been used extensively and very
successfully to determine the specific values of the critical exponents and their more general dependence on the range of the intermolecular forces
present \cite{LF2}. In 2D the energy cost of an interfacial configuration can be described by the mesoscopic continuum model
\begin{equation}
H[\ell]=\int dx\left(\frac{\Sigma}{2}\left(\frac{dl}{dx}\right)^2+V(\ell)\right)
\label{H}
\end{equation}
where $\ell(x)$ is a collective co-ordinate representing the local height of the liquid-gas  interface above the wall. Here $\Sigma$ is the stiffness
coefficient, equivalent to the tension $\sigma$ for isotropic fluid interfaces, while $V(\ell)$ is the binding potential which models the direct
interaction of the interface with the wall arising from intermolecular forces. The binding potential $V(\ell)$ can be thought of as describing the
underlying bare or mean-field wetting transition which would occur if the stiffness were infinite and interfacial fluctuation effects are suppressed.
To account for fluctuations it is necessary to evaluate the partition function for the model (\ref{H}). In 2D the scaling properties of the
interfacial roughness are insensitive to the choice of microscopic cut-off which is reflected by a universal (not depending on microscopic details)
relation between the roughness and the parallel correlation length. With an ``infinite momentum" cut-off the evaluation of the partition function $Z$
is then particularly straightforward since it is equivalent to a path integral and we can immediately write \cite{feynman, kleinert}
 \begin{equation}
Z(\ell,\ell';L)= \sum_n \psi^*_n(\ell)\psi_n(\ell') {\rm e}^{\beta E_n L}
 \end{equation}
where $\beta=1/k_B T$, $L$ is the lateral extent of the systems while $\ell,\ell'$ are the end point interfacial heights. Here $\psi_n$ and $E_n$ are
the eigenfunctions and eigenvalues of the continuum transfer matrix which takes the form of the Shr\"{o}dinger-like equation \cite{Lipkroll}
\begin{equation}
-\frac{1}{2\beta^2\Sigma}\psi_n''(\ell)+V(\ell)\psi_n(\ell)=E_n\psi_n(\ell)\,. \label{S}
\end{equation}
In the thermodynamic limit ($L\to\infty$) of an infinitely long wall the ground state identifies the singular contribution to the wall-gas surface
tension $\sigma_{\rm sing}=E_0$ and the probability distribution for the interface position follows as $P(\ell)=|\psi_0(\ell)|^2$. Similarly the
parallel correlation length describing the decay of the height-height correlation function along the wall is determined within the transfer-matrix
formulation as $\xi_\parallel=k_BT/(E_1-E_0)$.

The analysis of 2D wetting transitions using this transfer-matrix approach has already  been done in a great deal of detail by Kroll and Lipowsky
\cite{Lipkroll}. Suppose the bare wetting transition is continuous as described by the binding potential $V(\ell)=a\ell^{-p}+b\ell^{-q}+h\ell$ where
$q>p$ and the coefficient $a$ is considered negative at low temperatures. Provided that $b>0$ the condition $a=0$ (and $h=0$) represents the
mean-field critical wetting phase boundary \cite{Dietrich}. Solution of the Shr\"{o}dinger equation shows that the critical wetting transition falls
into several fluctuation regimes with the SFL regime, representative of short-ranged wetting holding for $p>2$. For $p<2$ we need only note that the
transition still occurs at the mean-field phase boundary $a=0$ although critical exponents are non-classical if $q>2$. However in the SFL regime, the
wetting temperature is lowered below its mean-field value since the interface is able to tunnel away from the potential well in $V(\ell)$ even though
$a<0$. Calculation shows that the singular part to the free-energy exhibits the anticipated scaling behaviour \cite{Fisher, Schick, Forgacs}
\begin{equation}
\sigma_{\rm sing}= t^{2}W(h |t|^{-3}) \label{1}
\end{equation}
identifying the universal values of the critical exponents $\alpha_s=0$ and $\Delta_s=3$ as quoted above.  Implicit here is that the scaling function
$W(x)$ is different below and above the wetting temperature and we have replaced $t$ with $|t|$ in the argument for convenience. The scaling of the
free-energy determines that the film thickness $\langle \ell\rangle\propto \frac{\partial \sigma_{\rm sing}}{\partial h }$ and correlation length
$\xi_\parallel^2 \propto \frac{\partial^2 \sigma_{\rm sing}}{\partial h^2 }$ must diverge as $\langle\ell\rangle\propto t^{-1}$ and
$\xi_\parallel\propto t^{-2}$  as $T\to T_w^-$ at bulk coexistence. These also follow from direct calculation. Indeed, the interfacial model
(\ref{H}) goes further and recovers precisely the scaling properties of energy density and magnetization correlation functions known from the exact
solution of the Ising model \cite{Ab,Burkhardt}. Above the wetting temperature the scaling of $\sigma_{\rm sing}$ also identifies, the correct
singular behaviour $\sigma_{\rm sing}\propto h^{2-\alpha_s^{co}}$ where $\alpha_s^{co}=4/3$ determines the singular contribution to the wall-gas
surface tension at the {\it{complete wetting}} transition occurring as $h\to 0$ \cite{AS,L}. Finally we mention that the case of binding potentials
which decay as an inverse square (i.e. $p=2$), referred to as the intermediate-fluctuation (IFL) regime, is marginal and the critical behaviour
subdivides into three further categories \cite{LN}.

In a related article Zia, Lipowsky and Kroll \cite{ZLK} also discussed what  happens if the binding potential $V(\ell)$ has a form pertaining to a
mean-field first-order wetting transition. Suppose that, at bulk coexistence, the potential has a long-ranged repulsive tail $V(\ell)= a\ell^{-p}$
(with $a>0$) which competes with a short-ranged attraction close to the wall. They showed that if $p>2$ the transition is continuous and belongs to
the SFL regime universality class of short-ranged critical wetting. In this regime fluctuation effects always cause the interface to tunnel through
the potential barrier in $V(\ell)$ when $T$ is sufficiently close to $T_w$. For $p<2$ the transition is first-order ($\alpha_s=1$) and the adsorption
diverges discontinuously at the wetting temperature. The latter follows from (\ref{S}) since at $T_w$ there is a zero energy bound state wavefunction
which determines that the probability distribution decays (ignoring unimportant constant factors) as $P(\ell)\propto exp(-\ell^{1-\frac{p}{2}})$.
Explicit results for $p=1$ confirm this for a restricted solid-on-solid model \cite{Privman}. The case $p=2$ is marginal but displays first-order
wetting with $\alpha_s=1$ for $a>3/8\beta^2\Sigma$ corresponding to sub-regime {\bf{C}} of the IFL regime \cite{LN}. In this case a zero energy bound
state wavefunction also exists at $T_w$ and determines that the probability distribution decays as $P(\ell)\propto \ell^{1-\sqrt{1+8\beta^2\Sigma
a}}$. This algebraic decay means, rather unusually, that not all moments of the distribution exist at $T_w$ \cite{LN}. Thus, for example, for
$1/\beta^2\Sigma > a >3/8 \beta^2\Sigma$ the adsorption diverges continuously as $T\to T_w$ even though the transition is strictly first-order. For
$3/8 \beta^2\Sigma>a>-1/\beta^2\Sigma$ the wetting transition is continuous with non-universal exponents (sub-regime {\bf{B}}) to which we shall
return shortly. Note that the parallel correlation length for all 2D first-order wetting transitions also diverges continuously with a universal
power-law $\xi_\parallel\sim t^{-1}$ independent of $p$. This is equivalent to the statement of hyperscaling, which also holds in the SFL regime,
since near $T_w$ the next wavefunction above the groundstate lies at the bottom of the scattering spectrum ($E_1=0$) and hence $\sigma_{\rm
sing}=-k_BT/\xi_\parallel$. This scenario is subtly different to first-order wetting in 3D where $\xi_\parallel$, as defined through the decay of the
height-height correlation function, remains finite as $T\to T_w^-$. However a continuously diverging parallel correlation length, very similar to
that occurring in 2D, can still be identified for 3D first-order wetting by considering the three phase region near a liquid droplet or alternatively
by approaching the wetting temperature $T_w$ from above along the prewetting line \cite{IR,I}.

\section{Apparent first-order behaviour in the SFL and IFL regimes}
One issue that has not been addressed concerns the size of the asymptotic critical region in either the  SFL regime or sub-regime {\bf{B}} of the IFL
regime when the interface has to tunnel through the potential barrier in $V(\ell)$. Let us consider the SFL regime first. For systems with
short-ranged forces and in zero bulk field, $h=0$, this can be modelled by the very simple potential
\begin{equation}
V(\ell)=-U\Theta(R-\ell)+c\delta(\ell-R)
\label{V}
\end{equation}
together with the usual hard-wall repulsion for $\ell<0$. Here $\Theta(x)$ is the Heaviside step function.  With $c\gg1$ this potential models the
competition between a short-ranged attraction (of depth $U>0$) and a large but also short-ranged repulsion similar to that arising in the Ising model
studies where the surface enhancement term competes with a surface field. We emphasise that precisely the same crossover scaling described below
emerges if we use a square-shoulder repulsion in place of the delta function. This choice of local binding potential is the simplest one that
incorporates a short ranged attraction and a repulsive potential barrier. It therefore has the same qualitative features as binding potentials
describing first-order wetting constructed from more microscopic continuum models \cite{Dietrich}. Here the coefficient $c$ is regarded simply as an
adjustable parameter in order to tune the size of the critical region but, more generally, will increase exponentially with the size and width of the
potential barrier. Without loss of generality we work in units where $R=1$ and also set $2\beta^2\Sigma=1$ for simplicity. Rather than vary the
temperature we equivalently decrease the depth of the attractive short-ranged contribution until the interface unbinds from the wall.  Elementary
solution of the Shr\"{o}dinger equation for the potential (4) determines that the ground state wavefunction behaves as $\psi_0(\ell)\propto \sin
(\sqrt{U+E_0}\ell)$ for $\ell<R$ and $\psi_0(\ell)\propto e^{-\sqrt{|E_0|}\ell}$ for $\ell>R$. The delta function contribution to the potential
necessitates that $\psi_0'(R^-)-\psi_0'(R^+)=c \psi_0(R)$ and continuity of the wavefunction immediately gives
\begin{equation}
-\sqrt{-E_0}-\sqrt{U+E_0}\cot\sqrt{U+E_0}=c\,.
\end{equation}
Therefore the wetting transition occurs when $U=U_w$ where $-\sqrt{U_w}\cot\sqrt{U_w}=c$.  For large $c\gg1$ the latter condition simplifies to
$U_w\approx c^2\pi^2/(1+c)^2$. Writing $U\equiv U_w+t$, it follows that if $t$ and $c^{-1}$ are small then the equation for the ground state energy
simplifies to
\begin{equation}
\sqrt{-E_0}\approx\frac{c^2}{2\pi^2}(E_0+t)
\label{quad}
\end{equation}
and solution of this quadratic equation determines that the singular part to the free-energy (recall that $\sigma_{\rm sing}=E_0$) behaves as
\begin{equation}
\sigma_{\rm sing}=-t_{Gi}\left(1-\sqrt{1+\frac{t}{t_{Gi}}}\right)^2
 \label{Cross}
\end{equation}
Here we have introduced a thermal {\it{Ginzburg}} scaling field $t_{Gi}=\pi^4/c^4$ which measures the  size of the asymptotic critical regime
\cite{RW}. For $t/t_{Gi}\ll 1$ the free-energy vanishes as $\sigma_{\rm sing}\approx-t^2/4t_{Gi}$ consistent with universal critical behaviour
characterising the SFL regime ($\alpha_s=0$). However for $t/t_{Gi}\gg 1$, that is outside the critical regime, the surface free-energy vanishes as
$\sigma_{\rm sing}\approx-t$ in accord with the expectations of a first-order phase transition. The expression (\ref{Cross}) has a form consistent
with phenomenological theories of crossover scaling $\sigma_{\rm sing}=-t A_{cr}(t/t_{Gi})$ with the scaling function behaving as $A_{cr}(x)\to 1$ as
$x\to\infty$ and $A_{cr}(x)\sim x/4$ as $x\to 0$. Similar crossover scaling has been used for interfacial delocalization transitions in 3D
\cite{Binder}. Two derivatives of $\sigma_{\rm sing}$ w.r.t $t$ determines that the surface specific heat or equivalently the surface susceptibility
behaves as
\begin{equation}
\chi_{11}\propto\frac{1}{t_{Gi}(1+\frac{t}{t_{Gi}})^{3/2}}
\end{equation}
which {\it{outside}} the critical regime, $\frac{t}{t_{Gi}}\gg 1$, shows the same apparent power-law $\chi_{11}\propto t^{-\gamma_{11}^{\rm eff}}$
with $\gamma_{11}^{\rm eff}=3/2$ seen in the Ising model studies \cite{Abraham2016}.

The present analysis can be generalised by considering tunnelling through a potential barrier in the IFL regime. This can be modelled by simply
adding a long-ranged term $a\ell^{-2}$, for $\ell>R$, to the potential $V(\ell)$ shown in (\ref{V}). Recall that for $a>3/4$ the wetting transition
is first-order while for $3/4>a>-1/4$ (and recall we have set $2\beta^2\Sigma=1$) it is continuous. This sub-regime {\bf{B}} is characterised by
strongly non-universal critical exponents with, for example, $2-\alpha_s=2/\sqrt{1+4a}$ from which all other exponents follows using hyperscaling etc
\cite{LN}. Setting $a=0$ recovers the results for the SFL regime described above. For completion we note that for $a<-1/4$ the interface is bound to
the wall with the condition $a=-1/4$ defining a line of wetting transition (sub-regime {\bf{A}} of the IFL regime \cite{LN}). These wetting
transitions, which display essential singularities, are no longer induced by variation of the short-ranged field $U$ and crossover scaling cannot be
considered. Within sub-regime {\bf{B}} the presence of the delta function repulsion at $\ell=R$ does not affect the asymptotic critical singularities
but once again significantly reduces the size of the asymptotic regime. In this case the wetting transition occurs when
$-\sqrt{U_w}\cot\sqrt{U_w}=c-\frac{1}{2}(1-\sqrt{1+4a})$ and writing $U=U_w+t$, it is straightforward to show that for small $t$ and small $c^{-1}$
the ground-state energy $E_0$ satisfies an equation similar to (\ref{quad}) but with the LHS replaced with $-E_0$ raised to the power
$(\sqrt{1+4a})/2$. In this way we can see that the crossover from first-order behaviour $\sigma_{\rm sing}\approx -t$ occurring for $t/t_{Gi}\gg 1$
to the asymptotic criticality $\sigma_{\rm sing}=-t_{Gi} (t/t_{Gi})^{2-\alpha_s}$, with $\alpha_s<1$, is described by the implicit equation (up to an
unimportant multiplicative constant)
\begin{equation}
\left(-\frac{\sigma_{\rm sing}}{t_{Gi}}\right)^{\frac{1}{2-\alpha_s}}=\frac{\sigma_{\rm sing}+t}{t_{Gi}} \label{gen}
\end{equation}
which recovers trivially (\ref{Cross}) when we set $\alpha_s=0$. This now shows the role played by the exponent  $\alpha_s$ in determining the
crossover from apparent first-order to critical wetting in two dimensions. In particular for {\it{fixed}} $t$, and in the limit $t_{Gi}\to 0$, this
has the expansion $\sigma_{\rm sing}=-t+{\it{O}}(t^{\frac{1}{2-\alpha_s}})$ where the coefficient of the singular correction term depends on
$t_{Gi}$. With $\alpha_s=0$ this is the same expansion of the free-energy, $\sigma_{\rm sing}=-t+{\it{O}}(\sqrt{t})$ found in the Ising model
calculations in the strong surface coupling limit; see in particular equations (15) and (17) of \cite{Abraham2016}. As noted by these authors it is
the presence of the non-analytic correction to the pure first-order singularity, $\sigma_{\rm sing}=-t$, which determines the apparent divergence of
the surface susceptibility and specific heat. It follows that, more generally, the value of the exponent $\gamma_{11}^{\rm eff}$ characterising the
apparent divergence of $\chi_{11}$ satisfies the exponent relation
\begin{equation}
(2-\gamma_{11}^{\rm eff})(2-\alpha_s)=1 \label{relation}
\end{equation}
Thus in sub-regime {\bf{B}} of the IFL regime, for $t/t_{Gi}\gg1$ the surface susceptibility would have a different apparent divergence
$\chi_{11}\propto t^{-\gamma_{11}^{\rm eff}}$ with a non-universal exponent
\begin{equation}
\gamma_{11}^{\rm eff}=2-\sqrt{\frac{1}{4}+2\beta^2a\Sigma}
\end{equation}
and we have reinstated the dependence on the stiffness coefficient $\Sigma$ for completion.  This recovers the Ising model result on setting $a=0$
corresponding to strictly short-ranged interactions. Note that as $a$ is increased towards the boundary with sub-regime {\bf{C}} the value of
$\gamma_{11}^{\rm eff}$ approaches unity. This means that exactly at the {\bf{B}}/{\bf{C}} regime border the apparent divergence of $\chi_{11}$,
occurring for $t/{t_{Gi}}\gg 1$, is near indistinguishable from the asymptotic divergence $\chi_{11}\propto 1/t (ln t)^2$ occurring as $t \to 0$
\cite{LN}. While the analysis described here applies only to systems with thermal interfacial wandering the exponent relation (\ref{relation} is
strongly suggestive that the same anomalous $3/2$ power-law divergence would be observed for apparent first-order wetting even in systems where the
interfacial unbinding is driven by quenched random-bond impurities since then the transition is also strictly second-order ($\alpha_s=0$)
\cite{Fisher,Forgacs,K}.

Returning to the case of short-ranged forces pertinent to the SFL regime we note that the expression (\ref{Cross}) also determines the apparent and
asymptotic divergences of the parallel correlation length. First note that the first excited state is bound to the wall ($E_1<0$) for $t>t_{NT}$ but
lies at the bottom of the scattering spectrum ($E_1=0$) for $t<t_{NT}$. Here $t_{NT}$ is the location of a non-thermodynamic singularity at which
$\xi_\parallel$ has a discontinuity in its derivative w.r.t $t$ similar to that reported in \cite{Upton}. For large $c\gg 1$ this occurs at
$t_{NT}\approx 3\pi^2$ far from the wetting transition and crossover scaling region. This means that for $t<t_{NT}$ the {\it{same}} hyperscaling or
rather hyperuniversal relation $\xi_\parallel=k_BT/|\sigma_{\rm sing}|$ applies equally inside ($t/t_{Gi}\ll1$) and outside ($t/t_{Gi}\gg 1$) the
asymptotic critical regime. Thus implies that the correlation length shows crossover between two different power-laws ; $\xi_\parallel\propto t^{-1}$
valid for $t/t_{Gi}\gg1$, characteristic of 2D first-order wetting, to $\xi_\parallel\propto t_{Gi}t^{-2}$ for $t/t_{Gi}\ll1$ describing the
asymptotic criticality of the SFL regime (2D second-order wetting).

\section{Rounded pre-wetting transitions for $T>T_w$}
Further insight into the crossover scaling behaviour in the SFL  regime can be seen off bulk-coexistence by adding a term $h\ell$ or $h(\ell-R)$ to
(\ref{V}). In this case for small $t$, $c^{-1}$ and $h$ the ground state energy is determined from solution of
\begin{equation}
-h^{\frac{1}{3}}\frac{Ai'(-E_0h^\frac{2}{3})}{Ai(-E_0h^\frac{2}{3})}\approx\frac{c^2}{2\pi^2}(E_0+t)
\label{FS}
\end{equation}
which, for $t>0$ recovers (\ref{quad}) when $h=0^+$. Here $Ai(x)$ is the Airy function which determines the decay of the wavefunction for $\ell>R$
\cite{AS,L}. It follows that the singular part of the free-energy scales as
\begin{equation}
\sigma_{\rm sing}= t W_{cr}\left(\frac{h}{|t|^\frac{3}{2}};t/t_{Gi}\right) \label{gen2}
\end{equation}
which is the more general result involving a crossover scaling function of  two variables and applies both above and below the wetting temperature.
In the asymptotic critical regime $t/t_{Gi}\ll 1$, the scaling function $W_{cr} (x;y)\to y W(xy^{-\frac{3}{2}})$ so that $\sigma_{\rm sing}=
-\frac{t^2}{t_{Gi}}W(h t_{Gi}^\frac{3}{2}/|t|^3)$. This is precisely the same scaling shown in (\ref{1}) but now including a dependence on $t_{Gi}$,
which recall determines the size of the asymptotic critical regime, appearing via metric factors. It follows that on approaching the wetting
transition, $T\to T_w^-$ at bulk coexistence, the adsorption ultimately diverges as  $\langle\ell\rangle\propto \sqrt{t_{Gi}}t^{-1}$ while for the
parallel correlation length we recover the expression $\xi_\parallel\propto t_{Gi}t^{-2}$ quoted above. These are the standard critical singularities
for the SFL regime but now reveal the dependence of the critical amplitudes on $t_{Gi}$. In particular the amplitude for the divergence of the
adsorption {\it{vanishes}} as $t_{Gi}\to 0$ equivalent to the adsorption jumping from a microscopic to macroscopic value. Note that the factors of
$t_{Gi}$ in $\sigma_{\rm sing}$, $\langle\ell\rangle$ and $\xi_\parallel$ are all consistent with the relation $\sigma_{\rm sing}\propto
-{\mathcal{A}}\sigma\langle\ell\rangle^2/\xi_\parallel^2$ where, within the SFL regime, $\mathcal{A}=8$ is a universal critical amplitude independent
of $t_{Gi}$. This is reminiscent of the ``bending energy" contribution to the free-energy in the heuristic scaling theory wetting transitions
\cite{LF2} and leads directly to the Rushbrooke equality $2-\alpha_s=2\nu_\parallel-2\beta_s$ discussed earlier.

The crossover scaling of $\sigma_{\rm sing}$ shown in (\ref{gen2}) depends on the scaling variable $h|t|^{-\frac{3}{2}}$ which is different to that
appearing in (\ref{1}) characteristic of the SFL regime. However this power-law dependence is in complete agreement with the predictions of the
phenomenological scaling theory of first-order wetting developed by Indekeu and Robledo \cite{IR,I}. Indeed setting $\alpha_s=1$ determines
$\nu_\parallel=1$ (from hyperscaling) and hence $\Delta_s=3/2$ (from $\Delta_s=3\nu_\parallel/2)$ all of which are consistent with the behaviour
found for $\sigma_{\rm sing}$ and $\xi_\parallel$ for $t/t_{Gi}\gg 1$. Note also that above the wetting temperature, and for $|t|/t_{Gi}\gg 1$, we
may approximate $\sigma_{\rm sing}\approx t W_{cr}(h|t|^{-\frac{3}{2}};-\infty)$. The value $3/2$ of the crossover (or equivalently the
Indekeu-Robledo first-order) gap exponent now determines that in the limit $h\to 0$ we recover the correct complete wetting singularity $\sigma_{\rm
sing}\propto h^\frac{2}{3}$ the amplitude of which must not depend on $t$. Thus the crossover scaling form (\ref{gen2}) provides a consistent link
between previous scaling theories of continuous and first-order wetting.

More explicitly, above the wetting transition and for $|t|/t_{Gi}\gg 1$, that is away from the immediate vicinity of $T_w$, the approximate solution
of (\ref{FS}) can be determined from simple expansion of the Airy function around its first zero. In this way it follows that the singular part to
the free-energy behaves as
\begin{equation}
\sigma_{\rm sing}\approx\frac{1}{2}\left(\lambda h^\frac{2}{3}+|t|-\sqrt{(\lambda h^\frac{2}{3}-|t|)^2+8ht_{Gi}^\frac{1}{2}}\right)
\end{equation}
where here $\lambda\approx 2.338$ is the negative of the first zero of the Airy function. If we could set $t_{Gi}=0$, which corresponds of course to
an artificial infinite potential barrier, then $\sigma_{\rm sing}=$Min$(|t|,\lambda h^\frac{2}{3})$. This determines a line of first-order phase
transition extending away from bulk coexistence located at $|t|=\lambda h^\frac{2}{3}$. For small $t_{Gi}$ these transitions are {\it{rounded}} on a
scale set by $h^\frac{1}{2}t_{Gi}^\frac{1}{4}$. Taking the derivative of $\sigma_{\rm sing}$ w.r.t $h$ determines that  $\langle\ell\rangle \approx
0$ for $|t|<\lambda h^\frac{2}{3}$ while $\langle\ell\rangle \approx h^{-\frac{1}{3}}$ for $t>\lambda h^\frac{2}{3}$. The sharp increase in the film
thickness therefore corresponds simply to a line of pseudo pre-wetting transitions. This line meets the bulk coexistence axis tangentially and the
power-law dependence on $h$ is in precise accord with the standard thermodynamic prediction for its location based on the Clapeyron equation
\cite{Dietrich}. Sitting at a given point along this line the parallel correlation length scales as $\xi_\parallel =t^{-1}\tilde\Lambda (t/t_{Gi})$
which follows from (\ref{gen2}) and also direct calculation of the spectral gap $E_1-E_0$. For $|t|/t_{Gi}\gg 1$ this reduces to $\xi_\parallel
=|t|^{-1}(|t|/t_{Gi})^{\frac{1}{4}}$ which is very large if $t_{Gi}$ is small. This lengthscale determines the rounding of the pre-wetting phase
transition equivalent to the characteristic size of the domains of the thick and thin prewetting states which are in pseudo phase coexistence. Moving
along the pre-wetting line away from the wetting temperature the lengthscale $\xi_\parallel$, and hence the size of the domains simply decreases,
indicating that the thin-thick transition is eventually smoothed away by fluctuations i.e. no pre-wetting critical point is encountered. On the hand
moving towards the wetting transition, while remaining along the pseudo pre-wetting line, the parallel correlation length eventually crossovers to
$\xi_\parallel\propto 1/|t|$. This is not indicative of any pseudo thin-thick phase coexistence but rather the usual thermal wandering of the
unbinding interface when $T_w$ is approaching along the thermodynamic path $h\propto |t|^\frac{3}{2}$. The above remarks are all consistent with the
general theory of the rounding of first-order phase transitions in pseudo one dimensional systems \cite{FP}

\section{Conclusions}
In this paper we have shown that recent Isings model studies which show apparent first-order wetting transitions are consistent with analysis of an
interfacial Hamiltonian model which also allows us to consider properties of the transition in the presence of marginal long-ranged forces and
occurring off bulk coexistence. Our study has revealed that the singular contribution to the surface free-energy shows a simple crossover scaling due
to the tunnelling of the interface through a potential barrier which generalises the standard scaling theory of critical wetting linking it
consistently with scaling predictions for first-order wetting. The form of the scaling function is explicitly calculated above and below the wetting
transition and illustrates the rounding of pseudo first-order phase transition in this low dimensional system. The crossover scaling occurring below
$T_w$, which is determined both within the SFL and IFL regimes, allows us to trace the value $3/2$ of the anomalous exponent $\gamma_{11}^{\rm eff}$
highlighted in the Ising model studies directly to the strict second-order nature of the critical wetting transition i.e. that $\alpha_s=0$. It would
be interesting to test the predicted non-universality of $\gamma_{11}^{\rm eff}$ in the IFL by adding a long-ranged external field to the Ising model
i.e. decaying as the inverse cube from the distance to the wall. Even for systems with short-ranged forces our predictions for the location of a
pseudo pre-wetting line above the wetting temperature can also be tested in numerical studies of the Ising model with a strong surface coupling
enhancement similar to that described in \cite{Abraham2016}. Finally we mention that similar apparent first-order behaviour and crossover scaling
should also occur in 2D for the interfacial delocalization transition near defect lines in the bulk if these too are now modified to include enhanced
couplings \cite{Forgacs, delocal}. Scenarios involving apparent first-order interfacial unbinding or delocalization in three dimensions are more
challenging.  However similar behaviour may occur at wedge filling transitions where fluctuation effects are enhanced compared to wetting and
interfacial tunnelling through a potential barrier can occur \cite{P1,P2}.

\begin{acknowledgments}
 \noindent This work was funded in part by the EPSRC UK grant EP/L020564/1, ``Multiscale Analysis of Complex Interfacial Phenomena''.
 A.M. acknowledges the support from the Czech Science Foundation, project 13-09914S.
\end{acknowledgments}

\end{document}